\documentclass[twocolumn,showpacs,aps,prb,amsmath,amssymb,floatfix]{revtex4}
\usepackage{graphicx}
\usepackage{dcolumn}
\usepackage{bm}
\begin{document}
\title{Effects of randomness on the critical temperature
in quasi-two-dimensional organic superconductors}
\author{Enver Nakhmedov$^{1,2}$, Oktay Alekperov$^{2}$ and Reinhold Oppermann$^{1}$}
\affiliation{ $^1$Institut f\"ur Theoretische Physik,
Universit\"at W\"urzburg, Am Hubland,
D-97074 W\"urzburg, Germany\\
$^2$Institute of Physics, Azerbaijan National Academy of
Sciences, H. Cavid ave. 33, AZ1143 Baku, Azerbaijan}
\date{\today}
\begin{abstract}
The effects of non-magnetic disorder on the critical temperature
$T_c$ of organic weak-linked layered superconductors with singlet in-plane pairing are considered.
A randomness in the interlayer Josephson coupling is shown to destroy phase
coherence between the layers and $T_c$ suppresses smoothly in a large 
extent of the disorder strength. Nevertheless the disorder of arbitrarily high 
strength can not  destroy completely the superconducting phase.
The obtained quasi-linear decrease of the critical temperature with increasing  
disorder strength is in good agreement with experimental measurements. 
\end{abstract}
\maketitle
\section{Introduction}

Organic molecular crystals $\kappa-(BEDT-TTF)_2X$ [abbreviated as $\kappa-(ET)_2X$] are
in the center of attention due to their unusual normal metallic and 
superconducting properties \cite{wsgc91}.
The flat $ET$ molecules in $\kappa-(ET)_2X$ organic metals  dimerize to form molecular units that
stack in planes on a triangular lattice \cite{iys06}. 
The anions $X$, which modify from 
$X=Cu[N(CN)_2]Cl$ through $X=Cu[N(CN)_2]Br$ to $X=Cu(NCS)_2$, separate the 
planes and accept one electron from each $BEDT-TTF$ dimer.
Most of the
ET-based superconductors (SCs) are strongly anisotropic quasi-two-dimensional (quasi-2D) conductors because
the conductivity is approximately isotropic within the layers of the ET donor molecules 
but smaller by a factor of $\sim 10^{3}$ in the perpendicular direction. 
Measurements of the superconducting coherence lengths \cite{kwok90} within- $\xi_{\|}$ and perpendicular 
$\xi_{\perp}$ to the superconducting planes in e.g. $\kappa-(ET)_2Cu[N(CN)_2]Br$ yield $\xi_{\|} \approx
37 \AA$ and $\xi_{\perp} \approx 4 \AA$, the latter of which is much smaller than the interlayer
distance $\sim 15 \AA$. This fact suggests that superconductivity in the direction perpendicular
to the plane may involve Josephson tunneling. 

Low temperature properties of organic SCs are known
to be very sensitive to disorder \cite{pk04}. Alloying with anions,
$x$-ray irradiation, or cooling rate controlled anion reorientation
introduces non-magnetic randomness into the system, however leaving
unchanged, to a large extent, in-plane molecular structures. 
Recently, the effects of non-magnetic disorder on superconductivity 
in organic $\kappa-(ET)_2Cu(SCN)_2$ have been studied
experimentally in Refs. \cite{aabo06, soyk11}. The non-magnetic 
disorder was introduced in these experiments via irradiation by either x-rays or protons 
\cite{aabo06,soyk11} and via partial substitution of $BEDT-TTF$ molecules with deuterated 
$BEDT-TTF$ or $BMDT-TTF$ molecules \cite{soyk11}. All disorder seems to affect the terminal ethylene
group and anion bound structures. The measurements for samples with molecular substitutions 
show \cite{soyk11} that the mean free path $l$ is longer than the in-plane coherent length $\xi_{\|}$,
indicating that the superconducting planes can be considered to be in clean limit.
$T_c$ was found \cite{aabo06} to fall quasi-linearly with defect density, and the dependence exhibits
a sharp change in slope from $0.31$ to $0.15$ around a threshold value of the interlayer residual resistivity
$\rho_0^{\ast}\approx 2~\Omega cm$. The main feature of the experiments is that the samples 
exhibit a superconducting ground state even at the highest defect densities, and there is not 
a SC-normal metal phase transition different from quasi-1D organic SCs \cite{no10},
where the randomness transforms the system into a normal metallic state. In the light of 
the experimental data, the Abrikosov-Gor'kov's theory \cite{ag58} for non-magnetic 
defects in non-s-wave SCs seems to fail to explain the experimental data.

We study in this article the 
effects of randomness in the Josephson coupling energy on the critical temperature
of weak-linked quasi-2D SCs. Therefore, the influence of a possible in-plane molecular 
disorder on the superconducting properties of the system is ignored.
Suppression of superconductivity in the presence of non-magnetic
impurities can in general be realized by destroying either
the modulus or the phase coherence of the order parameter \cite{nf98,ek95}.
Although strong fluctuations of the order parameter phase destroy off-diagonal
long-range order (ODLRO) in an isolated superconducting plane \cite{rice65}, the point-like
topological defects of a ``phase field'' such as ``vortex'' and ``antivortex'' of Kosterlitz and 
Thouless \cite{kt73} sets up a quasi-long range order in the system. 

\section{Classical phase fluctuation regime}

%

The strongly anisotropic organic SCs with in-plane singlet pairings are modeled 
as a regularly placed superconducting layers with Josephson-coupling 
between nearest-neighboring layers with a classical free energy functional
\begin{eqnarray}
&&\hspace{-5mm}F_{st}\{\varphi \} = N_s^{(2)}(T) \sum_j \int d^2r \bigg
\{\frac{\hbar ^2}{8 m_{\|}} \bigg[\bigg( \frac{\partial
\varphi_j}{\partial x} \bigg)^2 +\nonumber\\
&&\hspace{-5mm}+\bigg( \frac{\partial
\varphi_j}{\partial y} \bigg)^2\bigg]+\sum_{g=\pm 1}E_{j,j+g} \left[1 - \cos \big(
\varphi_j - \varphi_{j+g} \big)\right] \bigg \}, 
\label{freeenergy}
\end{eqnarray}
where $\varphi_j({\bf r})$ denotes the phase of the order
parameter $\Delta_j({\bf r}) = |\Delta_j| \exp (i
\varphi_j({\bf r}))$,  $N_s^{(2)}(T)$ is the
surface density of superconducting electrons; $N_s^{(2)}(T)= N_N^{(2)}(0)
\equiv N_N^{(2)} = \frac{p_F^2}{2 \pi \hbar^2}$ at $T \ge T_c^{(2)}$,
and $N_s^{(2)}(T) = N_N^{(2)}(0) \tau(T)$ with $\tau (T) =
\frac{T_c^{(2)} - T}{T_c^{(2)}}$ at $T \le T_c^{(2)}$.
The last term in Eq.(\ref{freeenergy}) describes the Josephson coupling with the energy $E_{j, j+g}$. 

Fluctuations of the order parameter modulus can be neglected for
pure SCs far from $T_c^{(2)}$, the mean-field critical temperature calculated for
an isolated layer. Therefore, the contributions to $F_{st}\{\varphi \}$ in Eq. (\ref{freeenergy}), 
coming from the modulus of the order parameter $| \Delta_{\bf j}|$, are omitted.

We assume the Josephson energy $E_{j, j+g}$ to be a random
parameter with Gaussian distribution, centered at the mean
value $E_g$, given by
\begin{equation}
P \{ E_{j, j+g} \} = (2 \pi W^2)^{-1/2} \exp \big \{ -
(E_{j, j+g} - E_{g})^2/(2 W^2) \big \}.
\label{gauss}
\end{equation}
$W^2$ is taken as a measure of disorder strength.

Employing the replica trick one can calculate the average value of
the order parameter $\cos (\varphi_{\bf j}(z))$
\begin{eqnarray}
\langle\langle\cos(\varphi_j({\bf r}))\rangle\rangle_{dis}= -\hspace{.1cm}T
\frac{\delta}{\delta \eta_j({\bf r})} \langle \ln \mathcal{Z}
\rangle |_{\eta_j({\bf z}) = 0}
\nonumber\\
= - T \hspace{.1cm}\frac{\delta}{\delta \eta_j ({\bf r})} \lim_{n \to 0}
\frac{ \partial \langle \mathcal{Z}^n \rangle}{\partial n}|_{\eta_j =
0}, \label{correlator}
\end{eqnarray}
where $\mathcal{Z} = \int \mathcal{D} \varphi \exp \big( -
\frac{1}{T} F_{st} \{ \varphi, \eta \} \big)$ is the
partition function with respect to the free energy functional $F_{st} \{
\varphi, \eta \}$ which contains, in addition to
Eq. (\ref{freeenergy}), the generating field term $\sum_j
\eta_j \cos ( \varphi_j({\bf r}))$. The double bracket
$\langle \langle \dots \rangle \rangle_{dis}$ is a shorthand notation
for the double average over thermodynamic fluctuations and
over disorder.

Integration out the Gaussian random variables yields
\begin{equation}
\langle \langle \cos ( \varphi_j({\bf r}))\rangle \rangle_{dis} =
\prod_{j, g}\int\frac{d \zeta_{j, g}}{\sqrt {2 \pi}} e^{ -
\frac{\zeta_{j, g}^2}{2}} \frac{\int \mathcal{D} \varphi \cos
(\varphi_j) e^{- \mathcal{F}/T}}{\int \mathcal{D} \varphi
e^{- \mathcal{F}/T}},
\label{av}
\end{equation}
with
\begin{eqnarray}
&&\hspace{-5mm}\mathcal{F} = N_s^{(2)} \sum_j \int d^2r \bigg\{
\frac{\hbar^2}{8 m_{\|}} \left[\left( \frac{\partial
\varphi_j}{\partial x}\right)^2 +\left( \frac{\partial
\varphi_j}{\partial y}\right)^2\right] +\nonumber\\ 
&&\hspace{-5mm}+\sum_g (E_g- W\zeta_{j,g}) [1 -
\cos(\varphi_j({\bf r}) -\varphi_{j+g}({\bf r}))]
\bigg\},
\label{F5}
\end{eqnarray}
where $\zeta_{j, g}({\bf r})$ denotes a Hubbard-Stratonovich auxiliary
decoupling field.
In order to clarify a character of the saddle-point for the variable $\zeta_{j,g}({\bf r})$,
one writes the expression (\ref{av}) for 
$\langle\langle \cos [\varphi_j({\bf r})-\varphi_{j+g}({\bf r})]\rangle \rangle_{dis}$ and find the
saddle-point $\zeta_{j_0,g_0}$ as
\begin{equation}
\zeta_{j_0,g_0}=\frac{W N_s^{(2)}}{T}\frac{\langle \cos[\varphi_{j_0}-\varphi_{j_0+g_0}]\rangle^2-
\langle \cos^2[\varphi_{j_0}-\varphi_{j_0+g_0}]\rangle}{\langle \cos[\varphi_{j_0}-\varphi_{j_0+g_0}]\rangle}
\label{saddle-point}
\end{equation}
The expression for the effective free-energy functional at 
the saddle-point $\zeta_{j_0,g_0}$, given by Eq. (\ref{saddle-point}), becomes similar to
that for a regular quasi-2D SC with renormalized inter-layer Josephson energy, 
$E_g \to E_g -\frac{W^2 N_s^{(2)}}{T}\frac{\langle \cos[\varphi_{j_0}-\varphi_{j_0+g_0}]\rangle^2-
\langle \cos^2[\varphi_{j_0}-\varphi_{j_0+g_0}]\rangle}
{\langle \cos[\varphi_{j_0}-\varphi_{j_0+g_0}]\rangle}$.  

\begin{figure}
\resizebox{.48\textwidth}{!}{%
\includegraphics[width=1cm]{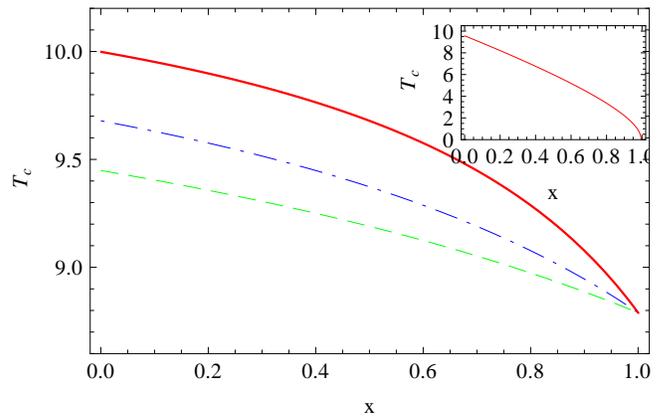}}
\caption {(Color online) Dependence of $T_c$ on $x=W^2/E_{\perp}^2$ for $q_0=0.6, 1.0$ and $2.0$ is depicted 
by dashed (green), dot-dashed (blue) and solid (red) curves, correspondingly. 
$T_c(q)$ increases with $q_0$ (or $E_{\perp}$), and reaches $T^{\ast}$ 
at highest value of the randomness $W=E_{\perp}$ when $x=1$. Insert shows $T_c(x)$ 
dependence according to the Abrikosov-Gor'kov's digamma function, which 
vanishes for higher values of $x$, instead of saturation in our case.}  
\label{Tc-classic}
\end{figure}

Equation for $T_c$ in quasi-2D SCs is derived from Eq. (\ref{av}) by
using the self-consistent mean-field method \cite{el74},
which consists in replacing the cosine term of Eq. (\ref{F5}) as
\begin{equation}
\sum_{\bf g} E_{\bf g} \cos(\varphi_{j}-\varphi_{j+g}) 
\to
\eta E_{\perp}  \langle \langle \cos(\varphi) \rangle
\rangle _{c}\cos \varphi (z),
\end{equation}
where $\eta$ is the coordination number, and $E_{\perp} \approx t_{\perp}^2/\epsilon_F$
with $t_{\perp}$ and $\epsilon_F$ being the interlayer tunneling integral and the Fermi energy. 
The phase correlations on the nearest-neighboring layers in this approximation are 
simplified by describing them as
a motion of a phason in the average field of phases with the most probable value, which coincides 
with the average value for a clean system $\langle \langle \cos (\varphi)\rangle  \rangle_c 
\equiv \langle \langle \cos (\varphi)\rangle \rangle_{dis}$ \cite{el74}. 
For a dirty system the most probable value differs strongly 
from the average value. Indeed, we assume that a distribution function of the order parameter 
in the presence of randomness is broad and asymmetric. This broadness and asymmetry 
becomes stronger around the critical temperature due to huge thermal fluctuations.
Therefore, knowledge of the arithmetic average is insufficient, and infinitely many
moments give a contribution to the distribution function of the order parameter at the tail. 
We identify $\langle \langle \cos(\varphi) \rangle
\rangle _{c}$ with the most probable or typical value of the order parameter. 
For the disordered SC we choose 
$\langle \langle \cos \varphi \rangle \rangle_c =\langle \langle \cos \varphi \rangle \rangle_{dis} -
\frac{\langle \langle \cos \varphi \rangle^2 \rangle_{dis} -
\langle \langle \cos \varphi \rangle \rangle_{dis}^2}{\langle \langle \cos \varphi \rangle \rangle_{dis}}$
which resembles a change made by the saddle-point (\ref{saddle-point}) in the free-energy functional.
The functional integral over the phases in Eq. (\ref{av}) can not
be evaluated yet, even after this simplification.
Taking advantage of the smallness of
$E_{\perp}\langle\langle\cos(\varphi)\rangle\rangle_c$
near $T_c$, however. an expansion of the integrand of Eq. (\ref{av})
in this quantity allows us to obtain the following equations for 
$\langle\langle\cos(\varphi)\rangle\rangle_{dis}$ and $\langle\langle\cos(\varphi)\rangle^2\rangle_{dis}$
\begin{equation}
\langle\langle\cos(\varphi)\rangle\rangle_{dis}=\frac{\eta N_s^{(2)}E_{\perp}}{k_BT} \int d^2 r 
\langle \cos \varphi (0) \cos \varphi ({\bf r}) \rangle_{0} \langle \cos \varphi \rangle_{c} 
\end{equation}
\begin{equation}
\langle\langle\cos(\varphi)\rangle^2\rangle_{dis}=\left(1+W^2/E_{\perp}^2\right) 
\langle \langle\cos \varphi \rangle \rangle_{dis}. 
\end{equation}
The final equation for $T_c$, obtained from the above written 
expressions, reads
\begin{equation}
\hspace{-3mm} 1 = \frac{\eta E_{\perp} N_s^{(2)}}{T_c}\biggl(1 - \frac{W^2}{ E^{2}_{\perp}}\biggr)
\int \langle \cos(\varphi({\bf r}))
\cos(\varphi(0))\rangle_0 d {\bf r}. 
\label{Tc}
\end{equation}
The phase-phase
correlator in Eq. (\ref{Tc}) is calculated in the clean limit of
the $2D$ free energy functional, obtained from
Eq. (\ref{freeenergy}) by setting $E_{\bf j,j+g} = 0$, which
yields, \cite{rice65,nf98}
\begin{eqnarray}
&&\langle \cos [\varphi ({\bf r})- \varphi (0)] \rangle_{0} =\nonumber\\
&&=\left\{ \begin{array}{ll} 
\left(\frac{\xi_{\|}}{r}\right)^{\frac{4k_BT}{\epsilon_F(1-T/T_c^{(2)})}}, & r > \xi_{\|}\\
\exp \left[-\frac{k_BT}{2 \epsilon_F(1-T/T_c^{(2)})}\left(\frac{r}{\xi_{\|}}\right)^2\right], & r<\xi_{\|}\\
\end{array} \right. 
\label{2Dcorrelator}
\end{eqnarray}
where $\xi_{\|} = \frac{\hbar \gamma v_F}{\pi^2 k_BT_c^{(2)}}$ with $\ln \gamma =c=0.577$ 
is the in-plane coherence length.
Real-space integration of the correlator (\ref{2Dcorrelator}) in Eq. (\ref{Tc}) 
for the critical temperature imposes the following restriction on the critical temperature
$-\frac{4 k_BT_c}{\epsilon_F(1-T_c/T_c^{(2)})}+2<0$ yielding $T_c>T^{\ast}$, where 
\begin{equation}
1/T^{\ast}=1/T_c^{(2)}+2 k_B/\epsilon_F. 
\label{T-ast}
\end{equation}
$T^{\ast}$ may be identified as
the Kosterlitz-Thouless transition temperature. The order parameter's phases correlation 
between nearest-neighbor layers disappears as $T_c$ approaches $T^{\ast}$, 
and system reveals effectively 2D superconducting behavior
at $T=T^{\ast}$. So, the critical temperature in quasi-2D SC varies in the interval of
$T^{\ast} < T_c<T_c^{(2)}$. One can estimate $T^{\ast}$ for $\kappa-(ET)_2Cu(SCN)_2$ 
SC with $T_c^{(2)} \approx 10.5 K$ for which the Fermi velocity and the effective mass
of electrons are measured \cite{soyk11,aabo06} to be  $v_F\approx 4 \times 10^4 m/s$ and 
$m^{\ast} \approx 3~m_0$, respectively, where $m_0$ is a free electron mass. These data yield $T^{\ast} \approx 8.8K$, 
which agrees well with maximally dropped critical temperature with disorder in 
Refs. \cite{aabo06,soyk11}.
The randomness in the Josephson energy in the presence of the order parameter 
phase fluctuations destroys the transverse stiffness in the system. 
Equations (\ref{Tc}) and (\ref{2Dcorrelator}) yield
\begin{equation}
1 = qt^2\left\{4(1-e^{-1/4t})+1/(1- t)\right\},
\label{Tc3}
\end{equation}
where $t$ is a dimensionless $T_c$-shift, $0 < t < 1$, introduced as
\begin{equation}
t = (\epsilon_F/2 k_B)
\left(1/T_c - 1/T_{c}^{(2)} \right),
\label{t-eq}
\end{equation}
and $q$ is a dimensionless parameter
\begin{equation}
q = \frac{4 \eta \gamma^2}{\pi^4}\left(\frac{t_{\perp}}{k_BT_c^{(2)}}
\right)^2\left(1-\frac{W^2}{E_{\perp}^2}\right) \equiv q_0 (1-x),
\label{q-eq}
\end{equation}
with $\ln \gamma = c=0.577$. $q$ decreases from its maximal value 
$q=q_0=\frac{4 \eta \gamma^2}{\pi^4}\left(\frac{t_{\perp}}{k_BT_c^{(2)}}\right)^2$ to zero 
as the disorder parameter $x=W^2/E_{\perp}^2$ increases from zero up to the maximal value $x=1$ for 
strong randomness $W \sim E_{\perp}$. The numeric solution of Eqs. (\ref{Tc3}) and (\ref{t-eq}) 
for the dependence of $T_c$ on $x$ is depicted in Fig. \ref{Tc-classic}.

Equation (\ref{Tc3}) is solved in two asymptotic limits. For $0< t <1/4$, which corresponds to
a weak disorder limit when $T_c$ varies around $T_c^{(2)}$, the exponential term is neglected,
yielding
\begin{equation}
\frac{1}{T_c}=\frac{1}{T^{\ast}}-\frac{40 \eta \gamma^2 E_{\perp}}{\pi^4 k_B(T_c^{(2)})^2}
\left(1-\frac{W^2}{E_{\perp}^2}\right)
\label{Tc-weak}
\end{equation}
In the limit $1/4<t<1$, which corresponds to relatively strong disorder limit when $T_c$ varies
around $T^{\ast}$, the exponential term in Eq. (\ref{Tc3}) is expanded yielding
\begin{equation}
\frac{1}{T_c}=\frac{1}{T_c^{(2)}}+\frac{2k_B}{\epsilon_F(1+q)}\approx \frac{1}{T^{\ast}}-
\frac{8 \eta \gamma^2 E_{\perp}}{\pi^4 k_B(T_c^{(2)})^2}
\left(1-\frac{W^2}{E_{\perp}^2}\right)
\end{equation}

\section{Quantum phase fluctuations regime}

The results, obtained above for the classical fluctuations are valid for weak randomness, when $T_c$
and $E_{\perp}$ have relatively high values. Quantum fluctuations have to be taken
into account for a strong disorder (small $T_c$ and $E_{\perp}$) limit.
The effects of randomness in the presence of quantum phase fluctuations can
be studied by starting from a Hamiltonian of weak-linked metallic layers with
in-layer attractive electron-electron interactions. Integrating out the electronic degrees
of freedom in the partition function, by following the method of Ambegaokar et al. \cite{aes82}, 
yields the following the expression for the dynamical free energy functional
\begin{equation}
F_{qu}\{\varphi\}=\frac{\hbar}{8V} \sum_{i,j}\int \int d{\bf r} d{\bf r'}d\tau K_{i,j}
\dot{\varphi}_i({\bf r},\tau)\dot{\varphi}_j({\bf r'},\tau) + F_{st}^{qu}\{\varphi \},
\end{equation} 
where the phases depend now on the imaginary ``time'' $\tau$ too, the dot on the phase 
means a ``time''-derivative, $K_{i,j}$ is the susceptibility \cite{aes82,nf98}. 
$F_{st}^{qu}\{\varphi\}$ is the stationary part of the dynamical free energy 
functional, which differs from the classical functional (\ref{freeenergy})
by additional integration over the imaginary ``time'' $\tau$. Repeating the procedure of derivation
of Eq. (\ref{Tc}) for the critical temperature in the case of the classical fluctuations, one arrives at
\begin{eqnarray}
&&1 = \eta E_{\perp} N_s^{(2)}(T)
\biggl(1 - \frac{W^2}{ E^{2}_{\perp}}\biggr)\times\nonumber\\
&&\times \int_0^{1/k_BT}d \tau \int d^2r \langle \cos[\varphi(0,0)]
\cos[\varphi({\bf r}, \tau)]\rangle_0. 
\label{Tc-qu}
\end{eqnarray}
 The phase-phase quantum correlator for a pure $2D$ SC is calculated  to yield
\begin{eqnarray}
&&\hspace{-5mm}\langle \cos [\varphi({\bf r}, \tau)-\varphi (0,0)]\rangle_0=\nonumber\\
&&\hspace{-5mm}=\exp\left\{-\frac{k_BT}{V}
\sum_{\omega_n}\sum_{k>0}\frac{2[1-\cos({\bf k \cdot r}-\omega_n\tau)]}
{\frac{\hbar^2N_s^{(2)}k^2}{4m^{\ast}}+\frac{\hbar^2}{4}\omega_n^2K(k)}\right\}
\end{eqnarray}
A straightforward calculation results in
\begin{eqnarray}
&&\hspace{-7mm} \langle \cos [\varphi ({\bf r}, \tau)- \varphi (0,0)] \rangle_0 =\nonumber\\
&&\hspace{-7mm} =\left\{ \begin{array}{ll} 
\exp\left[-\alpha + \frac{\alpha}{\sqrt{\left(\frac{2T\tau}{\beta}+1\right)^2+
\left(\frac{r}{\xi_{\|}}\right)^2}}\right], & \beta <1,\beta r<\xi_{\|}\\
(\beta r/\xi_{\|})^{-\alpha \beta}e^{-\alpha}, & \beta <1,~\beta r>\xi_{\|}\\
\exp\left[-\alpha\left(\frac{\beta r^2}{8\xi_{\|}^2}+\frac{T\tau}{\beta}\right)\right], 
& \beta >1,r<\xi_{\|}\\
(r/\xi_{\|})^{-\alpha \beta}, &\beta >1,~r>\xi_{\|}
\end{array} \right. 
\label{2Dcorrelator-qu}
\end{eqnarray}
where $\alpha=\frac{2}{\pi \xi_{\|} \hbar}\sqrt{\frac{m^{\ast}}{KN_s^{(2)}}}$ and
$\beta=\frac{2k_BT}{\hbar}\xi_{\|}\sqrt{\frac{m^{\ast}K}{N_s^{(2)}}}$ are the
dimensionless quantum parameters. Although $\alpha=\frac{4\pi k_BT_c^{(2)}}{\epsilon_F\tau^{1/2}}\alpha_0$
is proportional to the dynamical parameter 
$\alpha_0=\frac{1}{2\gamma}\left[\frac{\pi}{(2\hbar^2/m^{\ast})K}\right]^{1/2}$, 
which characterizes a quantum charging effect in the system, 
the other parameter $\beta=\frac{T}{\pi T_c^{(2)}\alpha_0 \tau^{1/2}}$ depends inversely on $\alpha_0$, 
nevertheless the product $\alpha \beta$  does not depend on $\alpha_0$. The quantum correlator 
(\ref{2Dcorrelator-qu})
becomes the classic one (\ref{2Dcorrelator}) for $\alpha_0 \to 0$ ($\alpha \ll 1$, $\beta \gg 1$, 
but $\alpha \beta \to const$).  
\begin{figure}
\resizebox{.48\textwidth}{!}{%
\includegraphics[width=1cm]{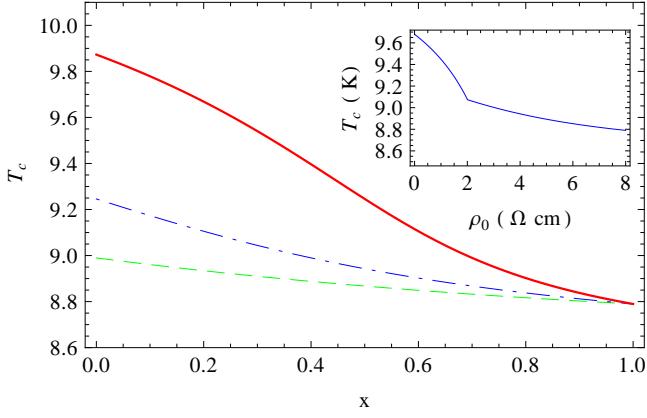}}
\caption{(Color online) Dependence of $T_c$ on $x=W^2/E_{\perp}^2$, according to Eq. (\ref{Tc-quantum}),  
for $a=0.5$ and three different values
of $q_0$: $q_0=0.6$ dashed (green) curve, $q_0=1.0$ dot-dashed (blue) curve and $q_0=2.0$ 
solid (red) curve. The dependence of $T_c$ on the residual resistivity $\rho_0$ in the insert is plotted by
combining $T_c(\rho_0)$ suppression in the classical fluctuations regime at $\rho_0 < \rho_0^{\ast} \simeq 2 \Omega cm$
with a slow suppression of $T_c(\rho_0)$ in the quantum phase fluctuations regime (corresponding to 
dot-dashed curves with $q_0=1$) at  $\rho_0 > \rho_0^{\ast}$.
}  
\label{SC-Tc-quantum}
\end{figure}
Equations (\ref{Tc-qu}) and (\ref{2Dcorrelator-qu}) yield a dependence of $T_c$ on $E_{\perp}$
and $W$ for two limiting cases $\beta >1$ and $\beta <1$. 
In both cases the integration of the correlator (\ref{2Dcorrelator-qu}) over coordinates 
imposes the restriction $T^{\ast}<T_c<T_c^{(2)}$ (or $\alpha \beta >2$), where $T^{\ast}$ 
is defined by expression (\ref{T-ast}). 
For $\beta <1$, which corresponds to strong charging regime, Eq. (\ref{Tc-qu}) after integration 
over ${\bf r}$ and $\tau$ results in non-linear equation for $t$ 
\begin{equation}
q\left\{\frac{t}{2a}+ a \left(1+\frac{t^3}{1-t}\right)\exp (-2\sqrt{a/t}~) \right\}=1,
\label{Tc-quantum}
\end{equation}
where, $a=\frac{2\pi^2 k_BT_c^{(2)}\alpha_0^2}{\epsilon_F}$ characterizes the
dynamic effects too, since $a \propto \alpha_0^2$. The numeric solution of Eq. (\ref{Tc-quantum}) is given 
in Fig. \ref{SC-Tc-quantum} for $a=1.2$ and different values of $q_0$. The quantum fluctuations reduce 
$T_c$ considerably and alter the results of the classical fluctuations regime at 
low temperatures and small $E_{\perp}$, which corresponds to strong
randomness or high resistivity in the $T_c(x)$ dependence. The slope of, e.g. the dashed 
(green) curve with $q_0=0.8$ changes from $0.54$ in the interval $0<x<0.5$ of 
Fig. \ref{Tc-classic} for the classic fluctuations regime to the value of $0.30$ in the interval $0.5<x<1$
of Fig. \ref{SC-Tc-quantum} for the quantum fluctuations regime, the ratio of which ($1.8$) is comparable 
with that ($\sim 2$) estimated for the experimental curve \cite{aabo06}.   
 
For $\beta > 1$ Eqs. (\ref{Tc-qu}) and (\ref{2Dcorrelator-qu}) are solved in the limit of 
$2<\alpha \beta <8$ yielding
\begin{equation}
T_c=T^{\ast} (1+q)/\{1+T^{\ast} q/T_c^{(2)}\}.
\end{equation}
The case of $\beta>1$ and $\alpha \beta > 8$ results in
\begin{equation}
1=qt^2\left\{4+ 1/(1-t)\right\},
\end{equation}
an approximate solution of which is given by Eq. (\ref{Tc-weak}). 
The case of $\beta>1$ or $\alpha <1$ corresponds to the weak quantum fluctuations limit, and, therefore, the results
do not depend on the dynamical parameter $\alpha_0$.

A detailed comparison of the results with the experiments needs to express $T_c$ on the interlayer residual
resistivity $\rho_0= \pi \hbar^4/(2e^2m^{\ast}a_{\perp}t_{\perp}^2\tau_t)$,  \cite{pk04} where $a_{\perp}$ is 
the interlayer distance. The inelastic scattering time $\hbar/\tau_t=\pi c_{imp}N(o)|U|^2$ is assumed to 
relate with a measure of the randomness $x \sim W^2$ as $W^2= \pi c_{imp}|U|^2/2$. 
It is necessary to take into account that 
the charging effect in the quantum fluctuations regime reduces the value of the interlayer tunneling integral $t_{\perp}$
(or the transverse rigidity) \cite{nf98}. Therefore, $\rho_0$ in the quantum fluctuations regime is 
rescaled for a given value of $x$ to a larger interval in comparision with that in the classical fluctuation regime.
The dependence of $T_c$ on $\rho_0$ is given in the insert of Fig. \ref{SC-Tc-quantum}.
 
\section{Conclusions}

In this paper we report disorder effects on $T_c$ of quasi-2D 
SCs with random Josephson coupling.
The interplay of non-magnetic disorder with quantum phase fluctuations
becomes a central factor in suppression of the superconducting phase
in organic quasi-2D SCs. A randomness in the 
interlayer coupling energy is shown to
decrease $T_c$ quasi-linearly, nevertheless the superconducting phase
does not completely vanish even at arbitrarily high strength of the disorder.
The present theory explains very well the recent experimental measurements given in 
Refs.\cite{aabo06,soyk11}. We neglect in this article
effects of in-plane disorder on $T_c$ in organic SCs. Such 
randomness results in suppression of $T_c$ due to the Anderson localization
for non-s-wave pairings, and it seems to destroy the homogeneity of the order parameter modulus leading to 
the formation of a cluster-like ``superconducting island'' inside the metallic phase.
On the other hand the in-plane disorder may ``pin'' the Kosterlitz-Thouless 
topological defects and destroy the quasi-long range order in the system. All these effects
deserve further investigation. 

\section{Acknowledgment}

This research was supported by the DFG under Grant No. Op28/8-1 and 
by the government of the Azerbaijan Republic under Grant No. EIF-2010-1(1)-40/01-22.

\end{document}